# COLLABORATION TOOLS AND PATTERNS
# FOR CREATIVE THINKING

Christian Kohls

Cologne University of Applied Sciences
Steinmüllerallee 1
Gummersbach, 51643, Germany
e-mail: christian.kohls@fh-koeln.de

## ABSTRACT

Many creativity methods follow similar structures and principles. Design Patterns capture such invariants of proven good practices and discuss why, when and how creative thinking methods match various situations of collaboration. Moreover patterns connect different forms with each other. Once we understand the underlying structures of creative thinking processes we can facilitate digital tools to support them. While such tools can foster the effective application of established methods and even change their properties, tools can also enable new patterns of collaboration.

## PATTERNS FOR CREATIVE THINKING

Many descriptions of creativity methods exist. What motivates the description as patterns is the generalization of similar methods, the reasoning for the actual form in terms of forces, and the contextualization and connection of the methods and tools. A design pattern captures similar solutions that fit to recurrent problems in specific contexts; they also explain how a solution addresses the conflicting forces and which consequences can be expected on applying the pattern. Very often a specific method implies the use of other methods. A pattern language captures such relations (Alexander, 1979). There are already some beautiful attempts to capture patterns about creative collaboration, presentation and learning in pattern languages (Iba, Ichikawa, Sakamoto & Yamazaki, 2011; Iba, Matsumoto & Harasawa, 2012; Iba & Isaku, 2013). On a high level of abstraction we can identify several general innovation processes. One pattern is to first understand the situation and goal, followed by rapid idea generation, postponed idea evaluation, and finally the implementation of ideas. Roger von Oech (2008) links these steps to different roles or perspectives: an *explorer* searches for information and resources, the *artist* turns them into new ideas, a *judge* evaluates the merits and drawbacks and the *warrior* brings ideas into action. Similar high level patterns exist in the various approaches to "design thinking". While the whole process can be seen as a high-level pattern, each phase can be implemented by different patterns on a more specific level (Kohls, 2014). For example the idea generation process can be stimulated by thought provocations, combination and variation of existing concepts, changes of perspective, random impulses, use of metaphors etc. Each of these patterns is an alternative to the higher level goal (such as idea generation or evaluation). On the other hand, each pattern describes the general structure of many similar methods. The "change of perspective" is a pattern that unfolds into more specific methods such as looking at the problem with the eyes of different stakeholders, thinking how superman or Sherlock Holmes would tackle the problem, reversing the goal, or imagining to be the solution. A benefit of pattern languages is that these relations can be made explicit. The context of a pattern describes when it could be used. For example "change of perspective" can be used to explore the problem field, find or evaluate ideas. It also suggests next steps such as evaluating the outcomes generated by this pattern. Another benefit of the pattern approach is that the description format reasons about why a specific method works. For example, "change of perspective" solves the problem that it is most likely to miss important facts, potential paths and undesired consequences if you look at things only from one direction. To understand all the potentials and liabilities, to see new paths and solutions, to give new meanings to a situation, to have original interpretations of an object, you should look at it from different angles. This reasoning is true for all the variants of "change of perspective". By identifying that "change of perspective" is a common pattern we can explain the psychological processes behind this structure for all methods of this kind. Moreover, once we understand the structure, we are able to develop new variants that fit to the current situation. Another example for many variations of the same pattern is a random impulse. It means to get an unbiased direction of thought by using random stimuli. There are many variations to achieve this, including word lists, randomly open a book or journal, using the first thing you see outside etc., yet the underlying principle is the same. The random impulse can be used with different other patterns together. One can use two or more random words and force a connection between them (the pattern of combining unrelated ideas), a random concept can be used as a metaphor for the current situation (e.g. after opening a page that shows a yacht, one can ask what parts of a boat are analogue to the current problem), or one can use the random impulse to take a new perspective (e.g. after seeing a postman outside one can ask how a postman would tackle the problem). This exemplifies another important principle of pattern languages. We can take one pattern and blend it with different other patterns and we will get an abundance of methods. Pattern descriptions can support this by pointing to related patterns. They can also point to useful tools for each pattern.

## DIGITAL TOOLS FOR CREATIVE THINKING

There are some important properties of digital tools that can support creative thinking patterns. Most patterns follow a sequence of steps and one can easily use wizards to guide the facilitation of methods. Templates for interactive whiteboards can trigger thoughts and encourage participants to use different perspectives. A digital device can also show patterns on different levels: "change of perspective" can be presented as a particular thought trigger ("What would superman do?") and with more details on demand (step-by-step instructions, examples, stimulating questions, reasoning etc.). Many apps with virtual card decks follow this principle. These creativity apps also let you draw random cards. In general digital tools are a great resource for randomness: with one click you can show random words, images or new combinations of attributes. Variation, combination and restructuring of parts and concepts is another advantage of interactive tools. Since all combinations of images, drawings, texts etc. can be saved at any time one is liberated to explore alternatives without losing results. By connecting devices and cloud services one can leverage collaborative innovation networks. Individuals can store their idea journals and observations online and share them with others to stimulate their thoughts, get feedback or work together on a problem.

## EXTREME COLLABORATION

A new pattern of collaboration emerges when individual and shared work spaces are connected for brainstorming sessions. Extreme Collaboration means parallel contribution of items from all participants to a shared space. Participants enter ideas, thoughts, or suggestions on their personal devices and simultaneously send them to a large (interactive) screen. They can also send photos, sketches and screenshots. This kind of collaboration can be hosted by using Twitter walls or shared whiteboards (e.g. ConceptBoard, Lino.it or Mural.ly). A special tool is XC Collaboration for SMART Boards. XC creates a random session and participants can join instantly to send ideas and photos to the interactive whiteboard. We have experimented with this approach in schools and observed new patterns of creative thinking that would not be possible without the tool. Since all participants write in parallel a huge quantity of items can be sent at the same time. Instead of adding ideas sequentially (as on traditional whiteboards), ideas are added rapidly in parallel. Hence, one can use more time to organize, follow-up and evaluate ideas. Moreover, since all students can write at the same time, it is no longer only the loudest who speak out. Every idea matters and there is no facilitator who filters any thoughts too early.

## OUTLOOK

Patterns can capture the variations and connections between several creativity methods and digital tools. Digital tools can change the ways in which thoughts are stimulated. In the next terms our department will evaluate existing apps and develop new apps as well. We have already seen that digital tools can enable novel methods. Extreme Collaboration shows new effects on innovation through parallel contribution and connecting individual ideas. Future work will consider the effectiveness of the method and explore the structuring of large number of contributions, i.e. hundreds or thousands of ideas sent in parallel).